\documentclass[%
 reprint,
 amsmath,amssymb,
 aps,
 prb,
]{revtex4-2}

\usepackage{graphicx,color}
\usepackage{dcolumn}
\usepackage{bm}
\usepackage{braket}

\begin{document}

\preprint{APS/123-QED}

\title{Delta-T noise in the Kondo regime}

\author{Masahiro Hasegawa}
\email{h.masahiro@rk.phys.keio.ac.jp}
\author{Keiji Saito}%
 \email{saitoh@rk.phys.keio.ac.jp}
\affiliation{Department of Physics, Keio university, Hiyoshi 3-14-1, Kohoku-ku, Yokohama, Japan}%

\date{\today}

\begin{abstract}
We study the delta-T noise in the Kondo regime, which implies the charge current noise under the temperature bias for the SU(2) Kondo quantum dot.
We propose an experimentally measurable quantity to quantify the low-temperature properties in the delta-T noise: $S_{\ell}=S(T_{\mathrm{L}},T_{\mathrm{R}}) - (1/2)[S(T_{\mathrm{L}},T_{\mathrm{L}}) + S(T_{\mathrm{R}},T_{\mathrm{R}})]$, which yields the shot noise expression in the noninteracting limit. We calculate this quantity for the SU(2) Kondo quantum dot in the particle-hole symmetric case.
We found that the $S_{\ell}$ exhibits qualitatively the same behavior in both the electrochemical potential biased case and the temperature biased case.
The quantitative difference appears as a difference of the coefficients of the noises, which reflects the difference of the Fermi distribution function:
electrochemical potential biased or temperature biased.
\end{abstract}

\maketitle

\section{Introduction}
Charge current noise has been intensively investigated both theoretically and experimentally. Current noise probes the mechanism of the charge transport in mesoscopic systems.
In particular, at zero temperature, the charge current noise induced by the voltage bias is described purely by the shot noise, by which we can observe the effective charge of carriers in mesoscopic conductors~\cite{Schottky1918,Blanter2000Jan}.
For example, it is known that the effective charge becomes $e/3$ for fractional quantum Hall systems~\cite{Saminadayar1997Sep,Picciotto1998Jun},  $2e$ for normal metal-superconductor junctions~\cite{Lefloch2003Feb}, and $((N+8)/(N+4))e$ for the SU($N$) Kondo quantum dot systems~\cite{Gogolin2006Jul,Sela2006Aug,Mora2008Jan,Mora2009Oct,Sakano2011Feb,Zarchin2008Jun,Yamauchi2011Apr,Ferrier2016Mar}.

\begin{figure}[t]
    \centering
    \includegraphics[width=0.75\linewidth]{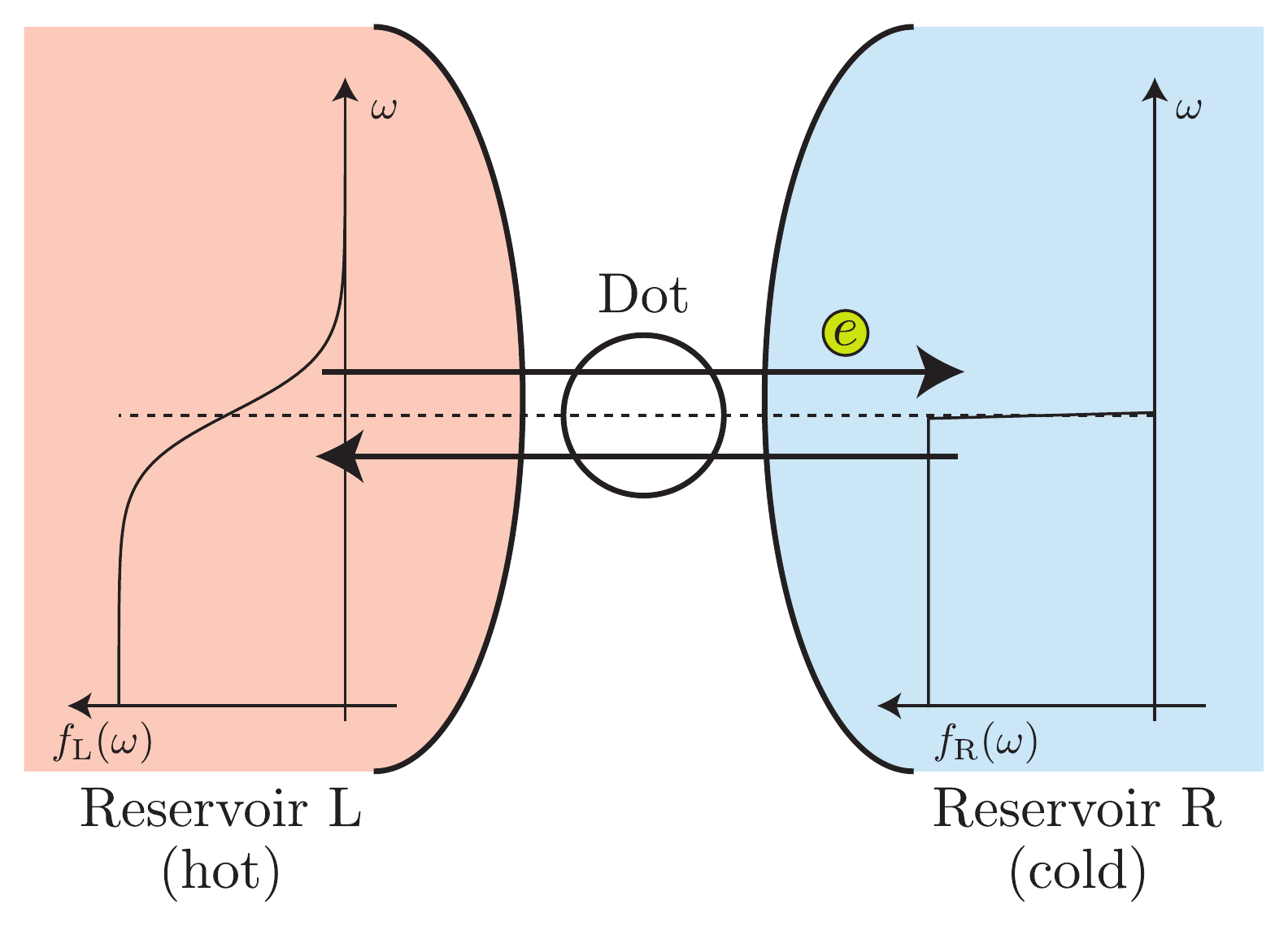}
    \caption{A schematic of the setup to measure the delta-T noise. A single-leveled quantum dot is connected to two electron reservoirs, reservoir L (hot reservoir) and reservoir R (cold reservoir).
      The curves shown in the red and blue areas stand for the Fermi distribution $f_{\alpha }(\omega) ~ (\alpha ={\mathrm L}, {\mathrm R})$. Electrons are transmitted from the reservoir L to R above the Fermi level and from R to L below the Fermi level.
    }
    \label{fig:model}
\end{figure}

Recently, owing to the development of measurement technologies in nanoscale conductors, it has become possible to measure the charge current noise due to the temperature bias, which is termed as {\it delta-T noise}.
In earlier works, the delta-T noise was proposed to detect the local noise in diffusive conductors ~\cite{Tikhonov2016Jul} and charge current noise in atomic-scale junctions~\cite{Lumbroso2018Oct}, electronic quantum circuits~\cite{Sivre2019Dec} and tunnel junctions~\cite{Larocque2020Feb}.
The delta-T noise is currently expected to be a new probe of quantum effects in charge transport which can not be observed by the shot noise measurement.
Unlike the shot noise, charges driven by temperature bias flow from each reservoir to the other (see Fig.~\ref{fig:model}).
A fair and open question is what type of information can be extracted from this new type of charge current noise.

We note that the fluctuation theorem can derive nontrivial relations between nonlinear transport coefficients even in the far-from-equilibrium regime~\cite{saitodhar,saitoutsumi,gaspard}.
However, the fluctuation theorem characterizes properties related to entropy production, while quantum nature in transport cannot be detected.
Hence, to figure out the open question above, model-dependent case studies are one of the critical directions.

In the recent paper~\cite{Rech2020Feb}, it was reported that the delta-T noise in the fractional quantum Hall system becomes negative compared with the noninteracting case.
This effect originates from the interaction effect, which cannot be observed in charge current noise under the voltage bias.
Stimulated by this intriguing indication, we discuss what types of interacting effects can appear in other important interacting transports.
In other words, we consider the following question: ``does the delta-T noise always exhibit this kind of intriguing effect in general quantum systems with strong interaction?''
To clarify this question, we study the delta-T noise via interacting quantum dots as the first case study.

\begin{table}[b]
    \begin{ruledtabular}
    \begin{tabular}{ccc}
         & Noninteracting limit ($U=0$) & Kondo limit ($U=\infty$) \\
        $C_{\mu}$ & $1/6$ & $5/3$ \\
      $C_T$ & $6\zeta(3) - 4\zeta(2)$ &  $(23/2) \zeta(3) + \left( 3 \ln 2 -8 \right) \zeta(2) $ \\
      & $(\sim 0.633)$ &  $(\sim 4.085)$ \\
    \end{tabular}
  \end{ruledtabular}
    \caption{ Summary of leading order contributions in the shot noise induced by bias voltage, $C_{\mu}$, and the delta-T noise induced by bias temperature, $C_T$. See Eqs. (\ref{eqn:curent_noise_RPT_mu}) and (\ref{rpttemp}) for the definitions of coefficients. \label{tab:table_noise}}
\end{table}

We consider the SU(2) Kondo quantum dot system in the particle-hole symmetric case. 
So far, the charge current noise under the voltage bias has been intensively studied theoretically~\cite{Gogolin2006Jul,Sela2006Aug,Mora2008Jan,Mora2009Oct,Sakano2011Feb} and experimentally~\cite{Zarchin2008Jun,Yamauchi2011Apr,Ferrier2016Mar}. It is known that the leading contribution of the noise starts with the cubic order of the voltage bias both in the noninteracting limit and in the Kondo limit. The difference appears in their coefficients, changing from $1/6$ in the noninteracting limit to $5/3$ in the Kondo limit (see Eqs. (\ref{eqn:curent_noise_RPT_mu}) and (\ref{eqn:curent_noise_coef_RPT_mu})). This difference appears due to the change of the charge carrier; while the charge transport is described by the free electrons in the noninteracting limit, it is described by the quasi-particle of the local Fermi liquid theory~\cite{Nozieres1974Oct} in the Kondo limit. For considering the delta-T noise, we employ the renormalized perturbation theory (RPT)~\cite{Hewson1993Jun,Hewson2001Oct,Oguri2001Sep} which reproduces the asymptotically exact results in the charge transport induced by bias voltage. Then, we consider the leading contribution in the delta-T noise. We show that the delta-T noise exhibits a similar structure to the noise under the voltage bias, i.e., the leading contribution starts with the cubic order of the temperature bias both in the noninteracting limit and in the Kondo limit, and their coefficients are modified by the interaction. The main results are listed in the Table \ref{tab:table_noise}.

This paper is organized as follows:
In Sec.~\ref{sec:Setup}, we introduce our model for Kondo quantum dots, the Anderson impurity model, and consider how to subtract the thermal noise contribution from the delta-T noise.
In Sec.~\ref{sec:Delta-T_noise_in_the_noninteracting_case}, the charge current noise in the noninteracting limit is discussed to compare the results in the Kondo limit.
In Sec.~\ref{sec:Delta-T_noise_in_the_Kondo_regime}, the charge current noise in the Kondo limit is discussed. Our main result is presented in this section.
In Sec.~\ref{sec:summary}, we summarize our results.
Detailed definitions and derivations are given in appendices.

\section{Setup \label{sec:Setup}}
\subsection{Model \label{sec:Model}}
To discuss the charge transport via Kondo quantum dots, we employ the Anderson impurity model (Fig. \ref{fig:model}), defined as
\begin{eqnarray}
    H = \sum_{r=\{\mathrm{L},\mathrm{R}\}} (H_r + H_{c,r}) + H_d ,
\end{eqnarray}
where
\begin{eqnarray}
    H_r &=& \sum_{k,s} \epsilon_{k} c_{rks}^{\dagger} c_{rks} , \\
    H_{c,r} &=& \sum_{k,s} \gamma (d^{\dagger}_s c_{rks} + c_{rks}^{\dagger} d_{s}) , \\
    H_d &=& \sum_{s} \epsilon_{d} d_{s}^{\dagger} d_{s} + U d^{\dagger}_{\uparrow} d_{\uparrow} d^{\dagger}_{\downarrow} d_{\downarrow} . 
\end{eqnarray}
$H_r$ is the Hamiltonian of the electron reservoir $r=\{\mathrm{L},\mathrm{R}\}$.
Here $c_{rks} (c_{rks}^{\dagger})$ is an annihilation (creation) operator of electrons with energy $\epsilon_k$ in the reservoir $r$. $k$ is the wavenumber and $s=\{ \uparrow, \downarrow \}$ is a spin index.
$H_d$ is the Hamiltonian of the quantum dot. $d_s (d_s^{\dagger})$ is an annihilation (creation) operator of electrons in the quantum dot with a spin $s$. $\epsilon_d$ is an energy level of the quantum dot and $U$ is the Coulomb interaction strength. $H_{c,r}$ is the Hamiltonian describing the symmetric tunnel coupling between the reservoir $r$ and the quantum dot with the coupling strength $\gamma$.
We prepare the reservoir L (R) in equilibrium with a temperature $T_{\mathrm{L}}$ $(T_{\mathrm{R}})$ and an electrochemical potential $\mu_{\mathrm{L}}$ $(\mu_{\mathrm{R}})$.
The Fermi distribution function of the reservoir $r$ is defined as $f_r(\omega) = [1+e^{(\omega-\mu_r)/T_r}]^{-1} $.
Throughout this paper, we assume that the density of states of the reservoirs are constant, known as the wide-band limit.
In this limit, the linewidth is defined as a constant, $\Gamma_{\mathrm{L}} = \Gamma_{\mathrm{R}} = \Gamma / 2 = 2 \pi \nu |\gamma|^2$, where $\nu$ is the density of states of the reservoirs.
The Lagrangian for this Hamiltonian is given as
\begin{eqnarray}
    &&\mathcal{L}(d_s,d_s^{\dagger},\epsilon_d,\Gamma,U) \nonumber \\
    &&= \sum_s d^{\dagger}_s(t) i \partial_t d_s(t) + \sum_{r,k,s} c_{rks}^{\dagger}(t) i\partial_t c_{rks}(t) - H(t) . \label{eqn:Lagrangian_original}
\end{eqnarray}
Throughout this paper, we assume $\hbar = k_B = 1$.

\subsection{Definition of the subtracted Delta-T noise \label{sec:Definition_of_the-subtracted_Delta-T_noise}}

The delta-T noise is the charge current noise induced by different temperatures in the electrodes with fixed electrochemical potential.
We consider the transport via the quantum dot depicted in Fig.\ref{fig:model}.
We define the charge current noise as
\begin{eqnarray}
    S = \int dt \ ( \Braket{ I(t) I (0)} - \Braket{I(t)} \Braket{I(0)} ) .
\end{eqnarray}
Here $I(t)$ is the symmetrized charge current defined as
\begin{eqnarray}
    I(t) &=& \frac{I_{\mathrm{L}}(t) - I_{\mathrm{R}}(t)}{2} , 
\end{eqnarray}
where $I_r(t)$ is the charge current at time $t$ flowing from the reservoir $r$.
In the particle-hole symmetric case, the difference of temperatures induces the nonequilibrium charge current noise, while average current is absent, because the transmission function is symmetric.

We first note that the current noise in the noninteracting case, which is especially denoted by $S_{\rm non}$, can be expressed as~\cite{Lesovik1989May,Lee1995Feb}
\begin{eqnarray}
  S_{\rm non}  &=& \frac{e^2}{\pi} \int d\omega \ \mathcal{T}(\omega) (1-\mathcal{T}(\omega)) (f_{\mathrm{L}}(\omega) - f_{\mathrm{R}}(\omega))^2 \nonumber \\
    & +& \frac{e^2}{\pi} \int d\omega \  \mathcal{T}(\omega) \sum_{r ={\mathrm L},{\mathrm R} } f_{r}(\omega) (1- f_{r}(\omega) )  \, ,  \label{eqn:noise_formal_noninteraction}
\end{eqnarray}
regardless of the parameter sets in the electrodes.
Here, $\mathcal{T}(\omega)$ is the transmission coefficient dependent on the frequency $\omega$.
In Eq. (\ref{eqn:noise_formal_noninteraction}), the first term can be regarded as the nonequilibrium contribution since this term never appear in the equilibrium situation, while the expression in the second line is identical to the thermal noise in equilibrium.
The main aim in this paper is to find interaction effects in the delta-T noise, comparing with the noninteracting case above.

To characterize nonequilibrium properties in delta-T noise, we need to note that contribution from the thermal noise, i.e., the Johnson-Nyquist noise, cannot be negligible.
Hence a sort of subtraction of the thermal noise contribution from the measured current noise is necessary to extract the nonequilibrium nature in the current noise. However, in general, identification of the thermal noise contribution in the nonequilibrium current noise is a difficult task.
Hence, we below {\it define} the subtracted delta-T noise with the two criteria. Our criteria on the subtraction in the delta-T noise are (i): {\it the subtraction must be experimentally feasible}, and (ii) {\it the subtracted noise must be reduced to the nonequilibrium contribution in the noninteracting limit, i.e., the first line in Eq. (\ref{eqn:noise_formal_noninteraction})}.

We consider the subtracted delta-T noise satisfying the criteria in two regimes: the small bias regime, $\Delta T /\bar{T} \ll 1$, and the large bias regime, $\Delta T /\bar{T} \sim {\mathcal O}(1)$.
Here, We denote the arithmetic mean value of the temperatures by $\bar{T}$, i.e., $\bar{T}={(T_\mathrm{L} + T_\mathrm{R}) /2}$, and denote the difference of temperatures by $\Delta T$, i.e., $\Delta T = |T_\mathrm{L} - T_\mathrm{R}|$.

The first regime, $\Delta T /\bar{T} \ll 1$, has been already discussed in Refs.~\cite{Lumbroso2018Oct,Rech2020Feb}.
The subtracted delta-T noise is defined as
\begin{eqnarray}
  S_{h}(T_\mathrm{L} ,T_\mathrm{R})  &:=& S(T_\mathrm{L}, T_\mathrm{R}) - S(\bar{T}, \bar{T}) \, .  \label{sh}
\end{eqnarray}
This definition provides the following leading order expansion for the noninteracting case~\cite{Lumbroso2018Oct} 
\begin{align}
 S_{h}(T_\mathrm{L} ,T_\mathrm{R}) &= k_{\rm B} {(\Delta T)^2 \over \bar{T}}\left( {\pi^2 \over 9} - {2\over 3}\right) G_0 \ \mathcal{T} (1- \mathcal{T} ) \, , \label{shexp}
\end{align}
where $G_0 $ is the quantum of conductance and $\mathcal{T}$ is the value of transmission coefficient at the Fermi energy~\footnote{Be sure that the frequency dependence of the transmission coefficient is ignored in this result.}. Obviously, this satisfies the two criteria (i) and (ii). Note that Ref.\cite{Lumbroso2018Oct} experimentally measured this quantity. The crucial difference from the shot noise induced by bias voltage lies in the prefactor in Eq. (\ref{shexp}), whose value is an indication that electrons can flow from both electrodes. Ref.~\cite{Rech2020Feb} shows the expansion for the fractional quantum Hall system that contains the interaction effects. Such expansion is possible for a relatively high temperature regime where the temperature difference is smaller than the average temperature. 

The second regime, $\Delta T /\bar{T} \sim {\mathcal O}(1)$, is relevant to our setup where we set $T_L=\Delta T$ and $T_R=0$ and consider the current noise in the Kondo regime. This case is relevant to the transport in the extreme low temperature regime. In this case, one needs to use another quantity as subtracted current noise. As one of simple definition, we here employ the following quantity  
\begin{align}
{S}_{\ell}(T_{\mathrm{L}  },T_{\mathrm{R}}) &:= S(T_{\mathrm{L}},T_{\mathrm{R}})- \frac{1}{2}(S(T_{\mathrm{L}},T_{\mathrm{L}})+S(T_{\mathrm{R}},T_{\mathrm{R}})) . \label{eqn:definition_tildeS}
\end{align}
This definition is experimentally measurable, hence it satisfies the criterion (i). In addition, from the exact expression (Eq.~(\ref{eqn:noise_formal_noninteraction})), the quantity $S_{\ell}$ satisfies the criterion (ii), because it yields the following expression for noninteracting case:
\begin{align}
  S_{\ell} (T_L , T_R)  &= \frac{e^2}{\pi} \int d\omega \ \mathcal{T}(\omega)(1-\mathcal{T}(\omega)) (f_{\mathrm{L}}(\omega) - f_{\mathrm{R}}(\omega))^2 \, . 
\end{align}
Obviously, $S_{\ell}$ satisfies the criterion (ii).
Hence, based on the quantity $S_{\ell}$, one can consider the interaction effect on the delta-T noise comparing the noninteracting limit. Notably, in the noninteracting transport case, one can easily check that the quantity $S_{\ell}$ reproduces the expansion of $S_{h}$, Eq.~(\ref{shexp}), once one takes high temperature regime limit.
Hence, one can expect that the quantity $S_{\ell}$ is available in a unified way, regardless of temperature regime.

In the subsequent sections, we discuss the subtracted delta-T noise using the expression $S_{\ell}$ for the parameters $T_L=\Delta T$ and $T_R = 0$. As a reference, we also discuss the shot noise that is a current noise induced by electrochemical potential bias at zero temperature, i.e., $\mu_{\mathrm{L}} = - \mu_{\mathrm{R}} = \Delta \mu / 2$ and $T_{\mathrm{L}} = T_{\mathrm{R}} = 0$.


%

\section{Delta-T noise and shot noise in  the noninteracting case \label{sec:Delta-T_noise_in_the_noninteracting_case}}

We first consider the non-interacting case with the particle-hole symmetry ($\epsilon_d = -U/2$ and hence $\epsilon_d =0$).
The charge current noise is expressed in Eq.  (\ref{eqn:noise_formal_noninteraction}). For the present Hamiltonian, the transmission coefficient is given by 
\begin{eqnarray}
    \mathcal{T}(\omega) = \frac{\Gamma^2/4}{(\omega-\epsilon_d)^2+\Gamma^2/4} .
\end{eqnarray}
For the parameter sets, $\mu_{\mathrm{L}} =  \mu_{\mathrm{R}} = 0$, $T_{\mathrm{L}} = \Delta T$, and $T_{\mathrm{R}} = 0$, the leading contribution of $S_{\ell} (\Delta T , 0)$ is calculated as
\begin{align}
    S_{\ell} (\Delta T , 0) &= \frac{2e^2}{\pi} \frac{(\Delta T)^3}{\Gamma^2} (6 \zeta(3) - 4 \zeta(2)) + {\mathcal O}((\Delta T)^5 ) . \label{eqn:noise_noninteracting_temperature} 
\end{align}
For comparison, we also present the shot noise expression induced by the electrochemical potential bias , by setting $\mu_{\mathrm{L}} = - \mu_{\mathrm{R}} = \Delta \mu / 2$ and $T_{\mathrm{L}} = T_{\mathrm{R}} = 0$. This is useful to understand the similarities and differences between the current noises induced by the temperature bias and the electrochemical potential bias.
For a zero temperature system, the thermal noise vanishes, and the charge current noise is described only by the shot noise.
The leading contribution of the noise is evaluated as 
\begin{eqnarray}
    S_{\rm shot} :=  \frac{2e^2}{\pi} \frac{(\Delta \mu)^3}{\Gamma^2} \frac{1}{6} + O((\Delta \mu)^5) . \label{eqn:noise_noninteracting_electrochemical}
\end{eqnarray}
A detailed derivation for (\ref{eqn:noise_noninteracting_temperature}) and (\ref{eqn:noise_noninteracting_electrochemical}) are presented in Appendix \ref{apx:Bias_parameter_expansion_of_the_shot_noise}.

Comparing the charge current noises under the temperature bias (\ref{eqn:noise_noninteracting_temperature}) and the electrochemical potential bias (\ref{eqn:noise_noninteracting_electrochemical}), we find that the leading order of the bias parameters both starts from the third order.
On the other hand, their coefficients are different; while a simple fraction appears in the electrochemical potential bias case, zeta functions appear in the temperature bias case, which originate from the Sommerfeld expansion.
This indicates that the difference in the non-interacting case just reflects how the bias parameters modify the Fermi distribution function; the electrochemical potential bias shifts the Fermi distribution function, whereas the temperature bias broadens it.


\section{Delta-T noise in the Kondo regime \label{sec:Delta-T_noise_in_the_Kondo_regime}}

Now, we discuss the charge current noise in the Kondo regime.
To discuss the Kondo problem for the particle-hole symmetric case ($\epsilon_d = - U/2$), the local Fermi liquid theory~\cite{Nozieres1974Oct} is a powerful analytical method.
The charge current noise in the Kondo limit ($U \to \infty$) has been discussed in terms of the phenomenological local Fermi liquid theory~\cite{Gogolin2006Jul,Sela2006Aug,Mora2008Jan}.
To discuss the crossover from the noninteracting limit to the Kondo limit, RPT~\cite{Hewson1993Jun,Hewson2001Oct,Oguri2001Sep} is a suitable method because it connects the results in the noninteracting regime with those in the Kondo regime in a microscopic picture.

\subsection{Renormalized perturbation theory}

In general, it is difficult to calculate the charge current and its noise in the interacting system because we cannot calculate the one-particle-irreducible (1PI) self-energy or the four-point full vertex function without any approximations.
However, in the Kondo problem, the electrochemical potentials and temperatures are assumed to be sufficiently smaller than the Kondo temperature.
In this case, one can discuss the physics only by considering the low-energy excitation around the Fermi level, or in other words, the lower-energy contributions of the 1PI self-energy and the four-point full vertex function.

RPT is the analytical method that renormalizes the lower-energy contribution of the 1PI self-energy and the four-point full vertex function into the parameters, $\epsilon_d$, $\Gamma$, and $U$, and reorganizes the perturbation scheme with the renormalized parameters, $\tilde{\epsilon}_d$, $\tilde{\Gamma}$, and $\tilde{U}$.
RPT has been used to calculate the transport coefficients in the Kondo regime, such as the charge and magnetic susceptibility~\cite{Hewson1993Jun}, charge conductance at low voltage bias~\cite{Oguri2001Sep}, current noise at low-voltage bias~\cite{Sakano2011Feb}, etc.

The advantage of RPT is that the results are {\it asymptotically exact} at small electrochemical potential and temperature bias.
Here, the phrase {\it asymptotically exact} means that one can obtain the exact result with the perturbation in finite order as long as one considers transport under small electrochemical potential and temperature bias.
For charge current and noise, the second order perturbation with respect to $\tilde{U}$ gives exact results up to the third order of the electro-chemical potential bias~\cite{Sakano2011Feb} and the temperature bias.

First, we discuss the renormalization of $\epsilon_d$ and $\Gamma$ by considering the one particle Green's functions (GFs).
The advanced (one particle) GF is defined as
\begin{eqnarray}
    G_s^A(\omega) = i \int dt \ e^{i\omega t} \Theta(t) \Braket{[d_s(t), d_s^{\dagger}(0)]_+}
\end{eqnarray}
Here $\Theta(t)$ is the Heaviside step function and $[\cdot,\cdot]_+$ is the anti-commutator.
The advanced Green's function (GF) in the interacting system for $\mu_{\mathrm{L}}=\mu_{\mathrm{R}}=0$ and $T_{\mathrm{L}}=T_{\mathrm{R}}=0$ is calculated as
\begin{eqnarray}
    G_s^A(\omega) = \frac{1}{\omega-\epsilon_d - i \Gamma /2 - \Sigma_{Us}^A(\omega)} .
\end{eqnarray}
Here, $\Sigma_{Us}^A(\omega)$ is the 1PI self-energy for the advanced GF induced by the Coulomb interaction.
This GF can be renormalized into the quasi-particle GF defined as
\begin{eqnarray}
    \tilde{G}_s^A(\omega) = z^{-1} G^A(\omega)= \frac{1}{\omega-\tilde{\epsilon}_d - i \tilde{\Gamma}/2 - \tilde{\Sigma}_{s}^A(\omega)} .
\end{eqnarray}
Here, $\tilde{\epsilon}_d$ and $\tilde{\Gamma}$ are the renormalized dot-level and linewidth, respectively, defined as
\begin{eqnarray}
    \tilde{\epsilon}_d &=& z(\epsilon_d + \Sigma_{Us}^A(0)) , \\
    \tilde{\Gamma} &=& z \Gamma ,
\end{eqnarray}
where $z=[1-\partial \left. \Sigma_{U,s}(\omega)/ \partial \omega \right|_{\omega=0} ]^{-1}$ is the wavefunction renormalization factor.
$\tilde{\Sigma}_{s}^A(\omega)$ is the 1PI self-energy induced by the renormalized interaction, defined as
\begin{eqnarray}
    \tilde{\Sigma}^A(\omega) = z \left[ \Sigma_{Us}^A(\omega) - \Sigma_{Us}^A(0) - \omega \left. \frac{\partial }{\partial \omega} \Sigma_{Us}^A(\omega) \right|_{\omega=0} \right] . \nonumber \\
\end{eqnarray}

Next, we consider the renormalization of $U$.
While the dot-level and the linewidth are renormalized by the one-particle dynamics, the Coulomb interaction is renormalized by the two-particle dynamics.
The renormalized Coulomb interaction is defined by the lower-energy part of the four-point full vertex function
\begin{eqnarray}
    \tilde{U} = (-i) z^2 \Gamma_{\uparrow \downarrow}^{++++}(0,0,0,0) .
\end{eqnarray}
Here $\Gamma_{\uparrow \downarrow}^{++++}(\omega_1,\omega_2,\omega_3,\omega_4)$ is the time-ordered component of the four-point full vertex function.
The superscript of the vertex function denotes the Keldysh indices.
Detailed definitions of GF and vetrx function is given in appendinx \ref{apx:ChargeCurrentNoiseGeneral}.

In RPT, one considers the perturbation theory with respect to the renormalized Coulomb interaction $\tilde{U}$.
The perturbation theory with the renormalized parameters is expected to be more accurate compared to that with the original parameters.
However, there is a problem that the Lagrangian with the renormalized parameters, $\mathcal{L}(\tilde{\epsilon}_d,\tilde{\Gamma},\tilde{U})$, is no longer equivalent to that with the original parameters, $\mathcal{L}(\epsilon_d,\Gamma,U)$.
This causes inconsistency between the results obtained by the renormalized parameters and those by the original parameters.
To maintain the consistency between the original and the renormalized models, one needs to introduce the counter-term Lagrangian 
\begin{eqnarray}
    && \mathcal{L}(d_s,d_s^{\dagger},\epsilon_d,\Gamma,U) \nonumber \\
    &&= \mathcal{L}(\tilde{d}_s,\tilde{d}_s^{\dagger},\tilde{\epsilon}_d,\tilde{\Gamma},\tilde{U}) + \mathcal{L}_{\mathrm{CT}}(\lambda_1,\lambda_2,\lambda_3) ,
\end{eqnarray}
where $\tilde{d}_s= z^{-\frac{1}{2}} d_s$ is the annihilation operator of the quasi-particle.
The counter-term Lagrangian is composed of three counter-terms:
\begin{eqnarray}
    \mathcal{L}_{\mathrm{CT}}(\lambda_1,\lambda_2,\lambda_3) &=&  \sum_s \tilde{d}_s^{\dagger}(t)(i\lambda_2 \partial_\tau - \lambda_1) \tilde{d}_s(t)  \nonumber \\
    &&\hspace{0.5cm}  - \lambda_3 \tilde{d}_{\uparrow}^{\dagger}(\tau) \tilde{d}_{\uparrow}(\tau) \tilde{d}_{\downarrow}^{\dagger}(\tau) \tilde{d}_{\downarrow}(\tau) .
\end{eqnarray}
The counter-terms, $\lambda_1$, $\lambda_2$, and $\lambda_3$, are defined as
\begin{eqnarray}
    \lambda_1 &=& -z \Sigma_{Us}^A (0) , \label{eqn:normalization_cond_1}\\
    \lambda_2 &=& z-1, \\
    \lambda_3 &=& z^2(U + i \Gamma_{\uparrow\downarrow}^{++++}(0,0,0,0)). \label{eqn:normalization_cond_3}
\end{eqnarray}
These three equations are called normalization conditions.
Another version of normalization conditions which are convenient for practical calculations are given in Appendix~\ref{apx:NormalizationConditions}.

At the end of this section, we should give a notice about the validity of RPT.
The renormalized parameters and the perturbation scheme are defined to reproduce the local Fermi liquid theory near the particle-hole symmetric point.
In other words, the RPT results are valid as long as the local Fermi liquid theory provides reliable results.
In terms of the charge current noise in the Kondo regime, it is known that the Fermi liquid theory is valid for the cubic order of the voltage bias~\cite{Gogolin2006Jul,Sela2006Aug,Mora2008Jan,Sakano2011Feb,Ferrier2016Mar}.
For the temperature bias case, the leading contribution is also the cubic order of the temperature bias.
This indicates that, analogous to the voltage bias case, the leading contribution in the temperature bias case can be obtained only by the lower-energy terms $\sim O(\omega^2,(\Delta \mu)^2,(\Delta T)^2)$ of the 1PI self-energies and the four-point full vertex functions, which is exactly calculated by the second-order perturbation of RPT~\cite{Oguri2001Sep,Sakano2011Feb}.

\subsection{Shot noise \label{sec:charge_current_noise_electochemical_interacting}}

First, we review the charge current noise under the electrochemical potential bias, $\mu_{\mathrm{L}} = - \mu_{\mathrm{R}} = \Delta \mu / 2$ and $T_{\mathrm{L}} = T_{\mathrm{R}} = 0$.
The charge current noise is calculated as
\begin{eqnarray}
    S_{\rm shot} = \frac{2e^2}{\pi} \frac{(\Delta \mu)^3}{\tilde{\Gamma}^2} C_{\mu} + O((\Delta \mu)^4), \label{eqn:curent_noise_RPT_mu}
\end{eqnarray}
where
\begin{eqnarray}
    C_{\mu} = \left[ \frac{1}{6} + \frac{3}{2} (R-1)^2 \right] . \label{eqn:curent_noise_coef_RPT_mu}
\end{eqnarray}
For a detailed derivation, see Appendix \ref{apx:ChargeCurrentNoiseGeneral} and \ref{apx:DerivationNoiseRPT}.
Here, $R$ is the Wilson ratio defined as
\begin{eqnarray}
    R = 1 + \frac{2}{1+\chi_c/\chi_s} = 1 + \frac{\tilde{U}}{\pi \tilde{\Gamma}},
\end{eqnarray}
where $\chi_c$ and $\chi_s$ are the charge and spin susceptibility, respectively.
This result is consistent with the result in Ref.~\cite{Sakano2011Feb}.
The difference from the noninteracting result is the $(R-1)^2$ term, which is the Fermi liquid correction.
Taking the Kondo limit ($U \to \infty$), it is known that the Wilson ratio converges to $2$; then one obtains the well-known value of the Fano factor~\cite{Gogolin2006Jul,Sela2006Aug,Mora2008Jan,Sakano2011Feb,Ferrier2016Mar},
\begin{eqnarray}
    \lim_{U\to\infty} \frac{S_{\rm shot}}{I_b} = \frac{5}{3}e ,
\end{eqnarray}
where $I_b$ is the backscattering current calculated as~\cite{Gogolin2006Jul,Sela2006Aug,Mora2008Jan,Sakano2011Feb}
\begin{eqnarray}
    I_b = \frac{2e}{\pi} \frac{(\Delta \mu)^3}{\tilde{\Gamma}^2} \frac{1+5(R-1)^2}{6} + O((\Delta \mu)^4 ) .
\end{eqnarray}

\subsection{Delta-T noise\label{sec:charge_current_noise_temperature_interacting}}

\begin{figure}
    \centering
    \includegraphics[width=\linewidth]{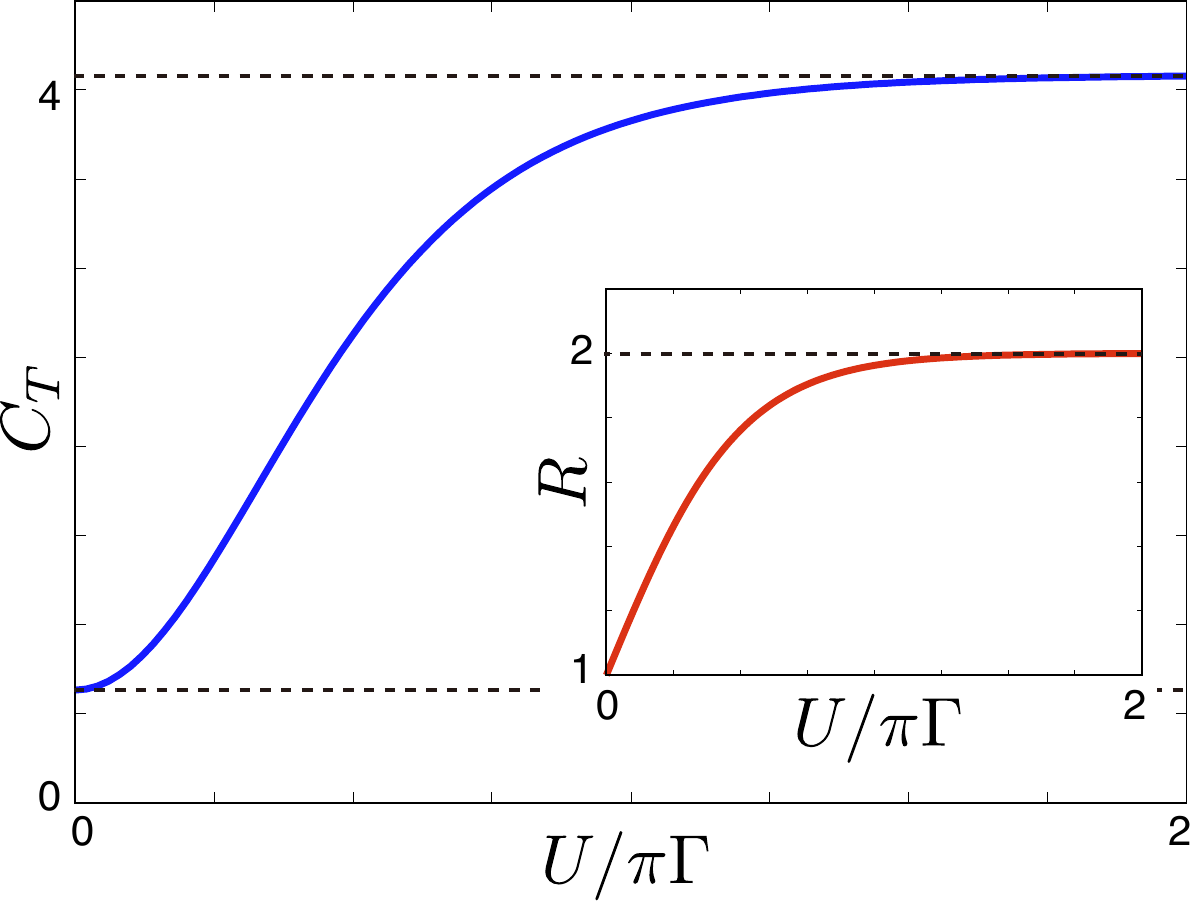}
    \caption{Interaction dependence of $C_T$.
    In the noninteracting limit ($U=0$), $C_T$ takes a positive value $6\zeta(3)-4\zeta(2) \sim 0.63$.
    As the interaction becomes stronger, $C_T$ monotonically increases and converges to the value, $(23/2) \zeta(3) + \left( 3 \ln 2 - 8 \right) \zeta (2)  \sim 4.08$, in the Kondo limit ($U \gg \pi \Gamma$).
    (Inset) Interaction dependence of the Wilson ratio $R$ calculated by the Bethe Ansatz~\cite{Wiegmann1983Sep,Kawakami1983Nov}.
    }
    \label{fig:noise_temp}
\end{figure}

Next, we consider the charge current noise in the Kondo region under temperature bias, $\mu_{\mathrm{L}} =  \mu_{\mathrm{R}} = 0$, $T_{\mathrm{L}} = \Delta T$, and $T_{\mathrm{R}} = 0$.
We consider that the temperature bias is sufficiently smaller than the Kondo temperature, $\Delta T \ll T_K$.
As with the noninteracting system, the thermal noise can not be ignored in the finite-temperature case; hence, we discuss the noise $S_{\ell}$.
Similar to the electrochemical potential bias case, charge current noise under temperature bias can be evaluated up to the cubic order as
\begin{align}
  S_{\ell}(\Delta T, 0) &= \frac{2e^2}{\pi}\frac{(\Delta T)^3}{\tilde{\Gamma}^2} C_T + O((\Delta T)^4) , \label{rpttemp}  
\end{align}
where
\begin{eqnarray}
    C_T &=& (6 \zeta(3) - 4\zeta(2))  \nonumber \\
    &&+ \left( \frac{11}{2} \zeta(3) + \left( 3 \ln 2  -4 \right) \zeta(2)  \right) (R-1)^2 . \label{eqn:curent_noise_coef_RPT_temp}
\end{eqnarray}
For a detailed derivation, see Appendix \ref{apx:ChargeCurrentNoiseGeneral} and \ref{apx:DerivationNoiseRPT}.
Figure \ref{fig:noise_temp} shows the interaction dependence of $C_T$.
In the noninteracting limit ($U=0$), it takes the value $6\zeta(3)-4\zeta(2)$.
In the Kondo limit ($U \to \infty$), it increases to the larger value, $ (23/2) \zeta(3) + \left( 3 \ln 2 - 8  \right) \zeta (2)$.
This is one of our main results.

Comparing the result in the temperature bias case (Eqs. (\ref{rpttemp}) and (\ref{eqn:curent_noise_coef_RPT_temp})) with that in the electrochemical potential case (Eqs. (\ref{eqn:curent_noise_RPT_mu}) and (\ref{eqn:curent_noise_coef_RPT_mu})), the interaction dependence is qualitatively the same in both cases; they appear as a square term of the Wilson ratio, $(R-1)^2$, and the coefficients, $C_{\mu}$ and $C_T$, increases as the interaction grows.
This indicates that Kondo physics acts on the charge current noise almost in the same manner even in the temperature bias case.

A difference appears in the coefficients;
the coefficient $C_T$ is described by the combination of the Riemann zeta functions, while the coefficient $C_{\mu}$ is described by simple fractions.
This difference reflects the difference of the Fermi distribution functions.
The charge transmission and collision process are described by the energy integral of the Fermi distribution functions.
The energy integrals of the Fermi distribution functions with electrochemical potential biases bring simple fractions and those with temperature biases bring the Riemann zeta functions.

\section{Connection to experiments}

\begin{figure}
    \centering
    \includegraphics[width=\linewidth]{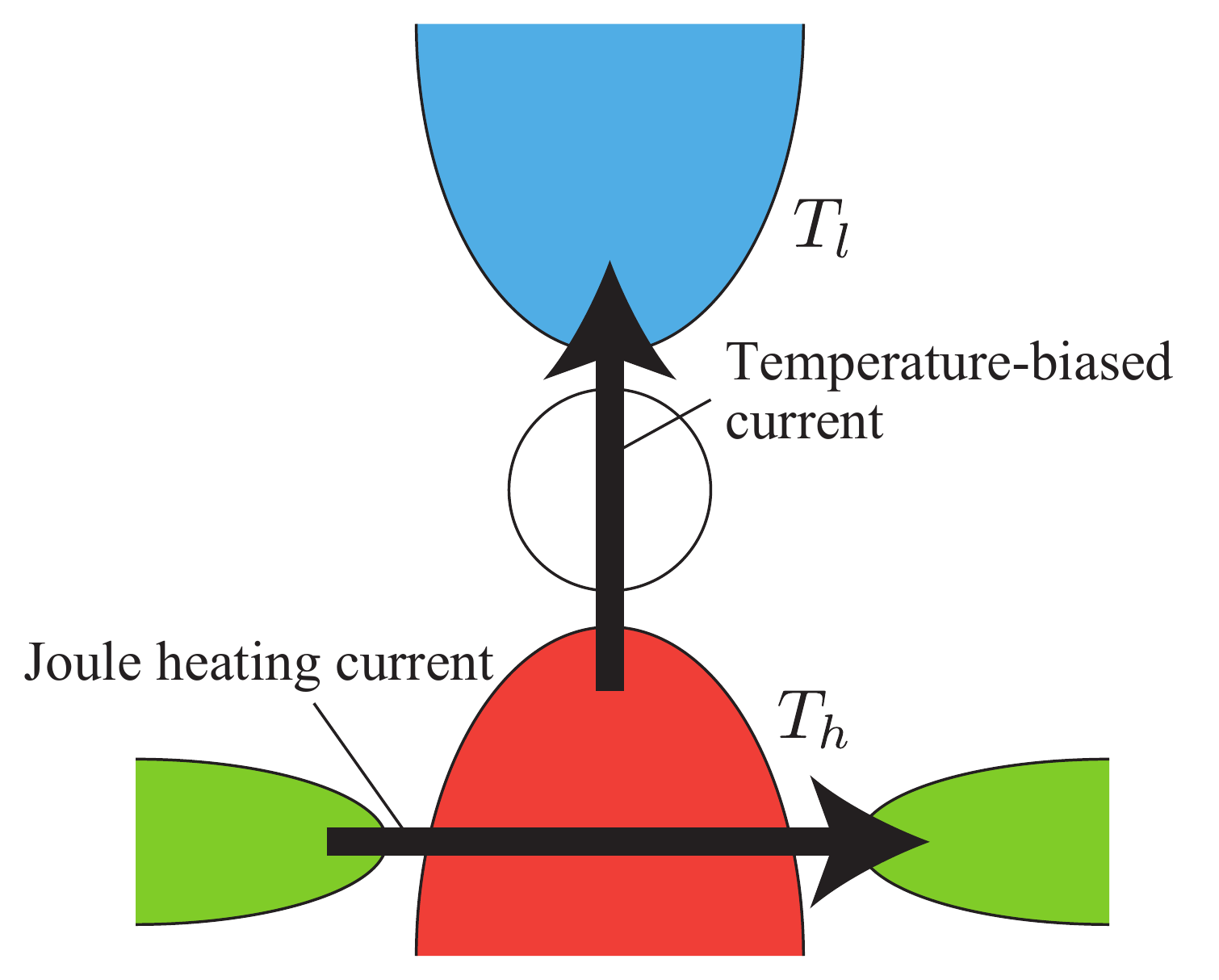}
    \caption{Schematic of the experimental setup that suits our setup.
    The red, blue, and green areas denote the source, drain, and ancilla electrodes, respectively.
    The voltage bias between the ancilla electrodes must be carefully tuned, so that no additional charge current flows between the source and drain electrodes due to the Joule heating current.
    }
    \label{fig:experiment}
\end{figure}

In this section, we briefly discuss the relationship between our theoretical results and the experiments.
Delta-T noise measurements have been realized via some experimental setups in previous studies~\cite{Lumbroso2018Oct,Sivre2019Dec,Larocque2020Feb}.
However, any experimental result that can discuss the delta-T noise in the Kondo regime has not been reported yet.

To confirm our results via experiments, it is required to build an experimental setup that can apply a temperature bias to the Kondo quantum dot.
The charge current noise measurement schemes in the Kondo quantum dot system have already been established by some studies~\cite{Zarchin2008Jun,Yamauchi2011Apr,Ferrier2016Mar}.
Thus, one can measure the delta-T noise by applying a temperature bias to the source and drain electrodes instead of a voltage bias.
The temperature difference that suits our setup can be realized in the following way (see Fig. \ref{fig:experiment}):
The entire system and the substrate are firstly prepared to the thermal equilibirum state with a temperature $T_l$.
Here $T_l$ is assumed to be almost zero as long as it is sufficiently smaller than the Kondo temperature.
Attaching two ancilla electrodes to the source electrode and applying charge current through the source electrode, the source electrode is heated up to a temperature $T_h > T_l$ due to Joule heating.
By carefully tuning of the voltage bias between the ancilla electrodes, a temperature bias between the source and drain electrodes can be realized without any voltage bias.
This Joule heating method is broadly employed in delta-T noise measurements~\cite{Lumbroso2018Oct,Sivre2019Dec,Larocque2020Feb}.
To measure the subtracted noise $S_{\ell}$, it is also necessary to measure the equilibrium noise in addition to the nonequilibrium noise.
The equilibrium noise with temperature $T_h$ can be measured by either changing the substrate temperature to $T_h$ or applying the same Joule heating current also to the drain electrode as well.

\section{Summary \label{sec:summary}}

We have discussed the charge current noise through a single-leveled quantum dot in the SU(2) Kondo region under temperature bias for the particle-hole symmetric case.
The charge current noise is composed of shot noise and thermal noise, which can not be measured separately in general.
We have discussed the noise $S_{\ell}$ defined in Eq. (\ref{eqn:definition_tildeS}) because it is the simplest definition of measurable noises, which can describe the far-from equilibrium noise beyond the Johnson-Nyquist noise and corresponds exactly with the shot noise in the noninteracting limit.
To discuss the Kondo effect, we have employed the renormalized perturbation theory and derived an analytical formula of $S_{\ell}$ under temperature bias by the second-order perturbation theory with respect to the renormalized interaction $\tilde{U}$.
This result is exact up to the order of $(\Delta T)^3$ for arbitrary strength of $U$.
This is one of our main results.

Next, we have compared the charge current noise in the temperature bias case with that in the electrochemical potential bias case.
The correction of the Kondo effect appears as a square term of the Wilson ratio $(R-1)^2$ in both temperature bias and electrochemical potential bias cases.
This indicates that the Kondo effect plays qualitatively the same role in both cases.
The difference appears as the difference of the coefficients of the noises.
The electrochemical biased one is described by simple fractions, while the temperature biased one is described by a combination of the Riemann zeta functions.
This difference originates from the difference of the Fermi distribuiton functions.

In this paper, we considered only the Kondo effect in the particle-hole symmetric case.
The charge transport via the Kondo quantum dot in the particle-hole asymmetric case has been an ongoing study~\cite{Filippone2018Apr,Oguri2018Mar,teratani2020Jan}.
Although the analytical method in these works is applicable only to voltage-biased transport, it would be able to treat a temperature-biased cases as well in the future.
The delta-T noise in the particle-hole asymmetric case remains a problem to be addressed in the future.
The heat current noises in the Kondo regime in electric systems and spin-boson systems \cite{Saito2013Nov} are also intriguing future subjects.

\begin{acknowledgments}
Authors thanks to the fruitful comments and discussions by R. Sakano and A. Oguri.
M.H. acknowledges financial support provided by the Grant-in-Aid for JSPS Fellows No. JP19J11360.
K.S. was supported by Grants-in-Aid for Scientific Research (JP16H02211, JP19H05603, JP19H05791).
\end{acknowledgments}

\appendix

\section{Bias parameter expansion of shot noise \label{apx:Bias_parameter_expansion_of_the_shot_noise}}
To consider $\Delta \mu$ and $\Delta T$ expansion of shot noise in the noninteracting system, we consider the following integral:
\begin{eqnarray}
    I = \int d\omega \ F(\omega) (f_{\mathrm{L}}(\omega) - f_{\mathrm{R}}(\omega))^2 , \label{eqn:integral_def}
\end{eqnarray}
where $F(\omega) = (1-\mathcal{T}(\omega)) \mathcal{T}(\omega)$.

For the electrochemical potential bias case, the difference between the Fermi distribution functions becomes a product of the Heaviside step functions
\begin{eqnarray}
    (f_{\mathrm{L}}(\omega) - f_{\mathrm{R}}(\omega))^2 = \Theta(\Delta \mu /2 - \omega) \Theta (\omega - \Delta \mu /2) .
\end{eqnarray}
Thus, the integral is evaluated as
\begin{eqnarray}
    I &=& \int_{-\Delta \mu /2}^{\Delta \mu} d\omega \ F(\omega) \nonumber \\
    &=& F(0) \Delta \mu + \frac{1}{24} \left. \frac{\partial^2 F(\omega)}{\partial \omega^2} \right|_{\omega=0} (\Delta \mu)^3  + O((\Delta \mu)^5)  . \nonumber \\
\end{eqnarray}
Substituting the equations
\begin{eqnarray}
    F(0) =  \left. \frac{\partial F(\omega)}{\partial \omega} \right|_{\omega=0} = 0 , \quad \left. \frac{\partial^2 F(\omega)}{\partial \omega^2} \right|_{\omega=0} = \frac{8}{\Gamma^2} , \label{eqn:detail_F}
\end{eqnarray}
one obtains Eq. (\ref{eqn:noise_noninteracting_electrochemical}).

For the temperature bias case, the difference between the Fermi distribution functions becomes
\begin{eqnarray}
    (f_{\mathrm{L}}(\omega) - f_{\mathrm{R}}(\omega))^2 &=& f_{\mathrm{L}}(\omega) + \Delta T \frac{\partial f_{\mathrm{L}}(\omega)}{\partial \omega} \nonumber \\
    && \hspace{1.0cm} - 2 f_{\mathrm{L}}(\omega) \Theta(-\omega) + \Theta(-\omega) . \nonumber \\ \label{eqn:fermi_dif_temp}
\end{eqnarray}
Substituting Eq. (\ref{eqn:fermi_dif_temp}) into Eq. (\ref{eqn:integral_def}) and using the following relations:
\begin{eqnarray}
    &&\int_{-\infty}^{0} d\omega \  F(\omega) f_{\mathrm{L}}(\omega) \nonumber \\
    &&= \int_{-\infty}^{0} d\omega \  F(\omega) - \ln 2 F(0) \Delta T + \frac{1}{2} \zeta(2) \left. \frac{\partial F(\omega)}{\partial \omega} \right|_{\omega=0} (\Delta T)^2\nonumber \\
    && \hspace{1.0cm}   -  \frac{3}{4} \zeta(3) \left. \frac{\partial^2 F(\omega)}{\partial \omega^2} \right|_{\omega=0} (\Delta T)^3 +O((\Delta T)^4) , \\
    &&\int_{0}^{\infty} d\omega \  F(\omega) f_{\mathrm{L}}(\omega) \nonumber \\
    &&=  \ln 2 F(0) \Delta T + \frac{1}{2} \zeta(2) \left. \frac{\partial F(\omega)}{\partial \omega} \right|_{\omega=0} (\Delta T)^2  \nonumber \\
    &&\hspace{1.0cm} +  \frac{3}{4} \zeta(3) \left. \frac{\partial^2 F(\omega)}{\partial \omega^2} \right|_{\omega=0} (\Delta T)^3  +O((\Delta T)^4) , 
\end{eqnarray}
then, the integral is calculated as
\begin{eqnarray}
    I &=& F(0) (2\ln2 - 1) \Delta T \nonumber \\
    && + \left( \frac{3}{2} \zeta(3) - \zeta(2) \right) \left. \frac{\partial^2 F(\omega)}{\partial \omega^2} \right|_{\omega=0} (\Delta T)^3  + O((\Delta T)^4) . \nonumber \\ \label{eqn:integral_temp}
\end{eqnarray}
Substituting Eq. (\ref{eqn:detail_F}) into Eq. (\ref{eqn:integral_temp}), one obtains Eq. (\ref{eqn:noise_noninteracting_temperature}).

\section{Charge current noise in the Keldysh formalism \label{apx:ChargeCurrentNoiseGeneral}}

\subsection{Green's functions}
To discuss the Kondo problem, one needs to consider the perturbation theory with respect to $U$; hence, the Keldysh formalism~\cite{Keldysh1965Oct} is the most suitable analytical tool because it can treat the perturbation theory in a systematic way.
In this section, we introduce the Keldysh Green's functions (GFs) for the Anderson impurity model.

The full GF of the electrons in the quantum dot is defined as
\begin{eqnarray}
    G_s^{\nu_1 \nu_2}(t_1,t_2) = (-i) \Braket{\mathcal{T}_K d_{s}(t_{1,\nu_1}) d_s^{\dagger}(t_{2,\nu_2}) } ,
\end{eqnarray}
where $\mathcal{T}_K$ is the time ordering operator on the Keldysh time contour.
$\nu_1$ and $\nu_2$ are the Keldysh indices for the time variables $t_1$ and $t_2$.
The Keldysh indices can take two values $+$ and $-$.
The $+$ index denotes that the time variable is on the forward path in the Keldysh time contour, and vice versa.
Here, the subscript $\cdot_{\nu_1}$ denotes that the time variable is on the $\nu_1$ path.
The full GF is calculated by the Dyson equation
\begin{eqnarray}
    G_s^{\nu_1 \nu_2}(t_1,t_2) &=& g_{s}^{\nu_1 \nu_2}(t_1,t_2) \nonumber \\
    && + \sum_{\nu_3,\nu_4} \int dt_3 dt_4  \ g_s^{\nu_1 \nu_3}(t_1,t_3) \nonumber \\
    && \hspace{2.0cm} \times \Sigma_{s}^{\nu_3\nu_4}(t_3,t_4) G_s^{\nu_4\nu_2}(t_4,t_2) , \nonumber \\ \label{eqn:dyson_equation_qpx}
\end{eqnarray}
where the summation of the Keldysh indices is defined as
\begin{eqnarray}
    \sum_{\nu} A^{\nu} =  A^{+} - A^{-} .
\end{eqnarray}
Here, $g_{s}^{\nu_1 \nu_2}(t_1,t_2)= \left. G_s^{\nu_1 \nu_2}(t_1,t_2) \right|_{\gamma=U=0}$ is the unperturbed GF and $\Sigma_{s}^{\nu_3\nu_4}(t_3,t_4)$ is the one-particle-irreducible (1PI) self-energy, defined as
\begin{eqnarray}
    \Sigma_{s}^{\nu_1\nu_2}(t_1,t_2) =  \Sigma_{Us}^{\nu_1\nu_2}(t_1,t_2) + \Sigma_{+s}^{\nu_1\nu_2}(t_1,t_2) , 
\end{eqnarray}
While $\Sigma_{Us}^{\nu_1\nu_2}(t_1,t_2)$ is the 1PI self-energy induced by the Coulomb interaction, $\Sigma_{+s}^{\nu_1\nu_2}(t_1,t_2)$ is that induced by the dot-reservoir coupling, defined as
\begin{eqnarray}
    \Sigma_{+s}^{\nu_1\nu_2}(t_1,t_2) = \gamma^2 \sum_{k} ( g_{\mathrm{L}ks}^{\nu_1\nu_2}(t_1,t_2) + g_{\mathrm{R}ks}^{\nu_1\nu_2}(t_1,t_2) ) , \nonumber \\
\end{eqnarray}
where $g_{rks}^{\nu_1\nu_2}(t_1,t_2)$ is the unperturbed GF of the electrons in the reservoir $r$ defined as
\begin{eqnarray}
    g_{rks}^{\nu_1\nu_2}(t_1,t_2)= (-i) \Braket{\mathcal{T}_K c_{rks}(t_{1,\nu_1}) c_{rks}^{\dagger}(t_{2,\nu_2})}_{\gamma=0} . \nonumber \\
\end{eqnarray}

The Dyson equation (Eq.~(\ref{eqn:dyson_equation_qpx})) is not useful to compute because all the Keldysh components of GFs are integrated with each other.
To avoid this problem, it is useful to use the advanced and retarded components of GF defined as
\begin{eqnarray}
    G_s^A(t_1,t_2) = G_s^{++}(t_1,t_2) - G_s^{-+}(t_1,t_2) , \\
    G_s^R(t_1,t_2) = G_s^{++}(t_1,t_2) - G_s^{+-}(t_1,t_2) . 
\end{eqnarray}
The Dyson equation for the advanced (retarded) GF is given as
\begin{eqnarray}
    G_s^{A(R)}(t_1,t_2) &=& g_{s}^{A(R)}(t_1,t_2) \nonumber \\
    && + \int dt_3 dt_4  \ g_s^{A(R)}(t_1,t_3) \nonumber \\
    && \hspace{2.0cm} \times \Sigma_{s}^{A(R)}(t_3,t_4) G_s^{A(R)}(t_4,t_2) . \nonumber \\
\end{eqnarray}

\subsection{Two-particle Green's functions}

For interacting systems, it is necessary to consider the two-particle collision due to the Coulomb interaction within the quantum dots, that is described in the two-particle GF in Keldysh formalism.
The two-particle GF is defined as
\begin{eqnarray}
    &&D_{s_1s_2}^{\nu_1\nu_2\nu_3\nu_4}(t_1,t_2,t_3,t_4)  \nonumber \\
    &&= (-i)^2 \Braket{\mathcal{T}_K d_{s_1}(t_{1,\nu_1}) d_{s_2}(t_{3,\nu_3}) d_{s_2}^{\dagger}(t_{4,\nu_4}) d_{s_1}^{\dagger}(t_{2,\nu_2})} \nonumber \\
    &&\hspace{3.0cm} - G_{s_1}^{\nu_1\nu_2}(t_1,t_2) G_{s_2}^{\nu_3\nu_4}(t_3,t_4) .
\end{eqnarray}
Analogous to the Dyson equation, the two-particle GF can be calculated by the four-point full vertex function $\Gamma_{s_1s_2}^{\nu_5\nu_6\nu_7\nu_8}(t_5,t_6,t_7,t_8)$ as
\begin{eqnarray}
    &&D_{s_1s_2}^{\nu_1\nu_2\nu_3\nu_4}(t_1,t_2,t_3,t_4) \nonumber \\
    &&= - \delta_{s_1,s_2} G_{s_1}^{\nu_1\nu_4}(t_1,t_4) G_{s_1}^{\nu_3\nu_2}(t_3,t_2) \nonumber \\
    &&\hspace{0.5cm} + \sum_{\nu_5, \cdots, \nu_8} \int dt_5 \cdots dt_8 \  \Bigl[  G_{s_1}^{\nu_1\nu_5}(t_1,t_5)G_{s_1}^{\nu_6\nu_2}(t_6,t_2)  \nonumber \\
    &&\hspace{4.0cm} \times G_{s_2}^{\nu_3\nu_7}(t_3,t_7) G_{s_2}^{\nu_8\nu_4}(t_8,t_4) \nonumber \\
    &&\hspace{4.0cm} \times \Gamma_{s_1s_2}^{\nu_5\nu_6\nu_7\nu_8}(t_5,t_6,t_7,t_8) \Bigr] , \nonumber \\
\end{eqnarray}
where $\delta_{s_1,s_2}$ is the Kronecker delta.

\subsection{Charge current and noise \label{sec:charge_current_and_noise}}
The steady charge current from the reservoir $r$ is defined as
\begin{eqnarray}
    \Braket{I_r(t)} &=& - e \frac{d}{dt} \sum_{k,s} \Braket{c_{rks}^{\dagger}(t) c_{rks}(t)} 
    \nonumber \\
    &=& e (-i) \sum_{k,s} \Braket{ \gamma d^{\dagger}_s(t) c_{rks}(t) - \gamma^{\ast} c_{rks}^{\dagger}(t)d_{s}(t) } . \nonumber \\ \label{eqn:current_definition}
\end{eqnarray}
For the symmetric dot-reservoir coupling system, one can define the symmetrized steady charge current as
\begin{equation}
    \Braket{I(t)} = \frac{\Braket{I_L(t)}-\Braket{I_R(t)}}{2} .
\end{equation}
The symmetrized steady charge current can be calculated for arbitrary $U$ by the Meir-Wingreen formula~\cite{Meir1992Apr}:
\begin{eqnarray}
    \Braket{I(t)} = \frac{e}{\pi} \int d\omega \  \mathcal{T}(\omega) (f_{\mathrm{L}}(\omega)-f_{\mathrm{R}}(\omega)) .
\end{eqnarray}
Here, $\mathcal{T}(\omega)$ is the transmission function for the interacting system, defined as
\begin{eqnarray}
    \mathcal{T}(\omega) = \frac{\Gamma}{2} \mathrm{Im} [ G_s^A(\omega)] ,
\end{eqnarray}
where $G_s^A(\omega)=G_s^{++}(\omega)-G_s^{-+}(\omega)$ is the advanced GF.
Here, we drop the subscript $s$ of the transmission function because it is spin independent for the no magnetic field case.

\begin{figure}
    \centering
    \includegraphics[width=\linewidth]{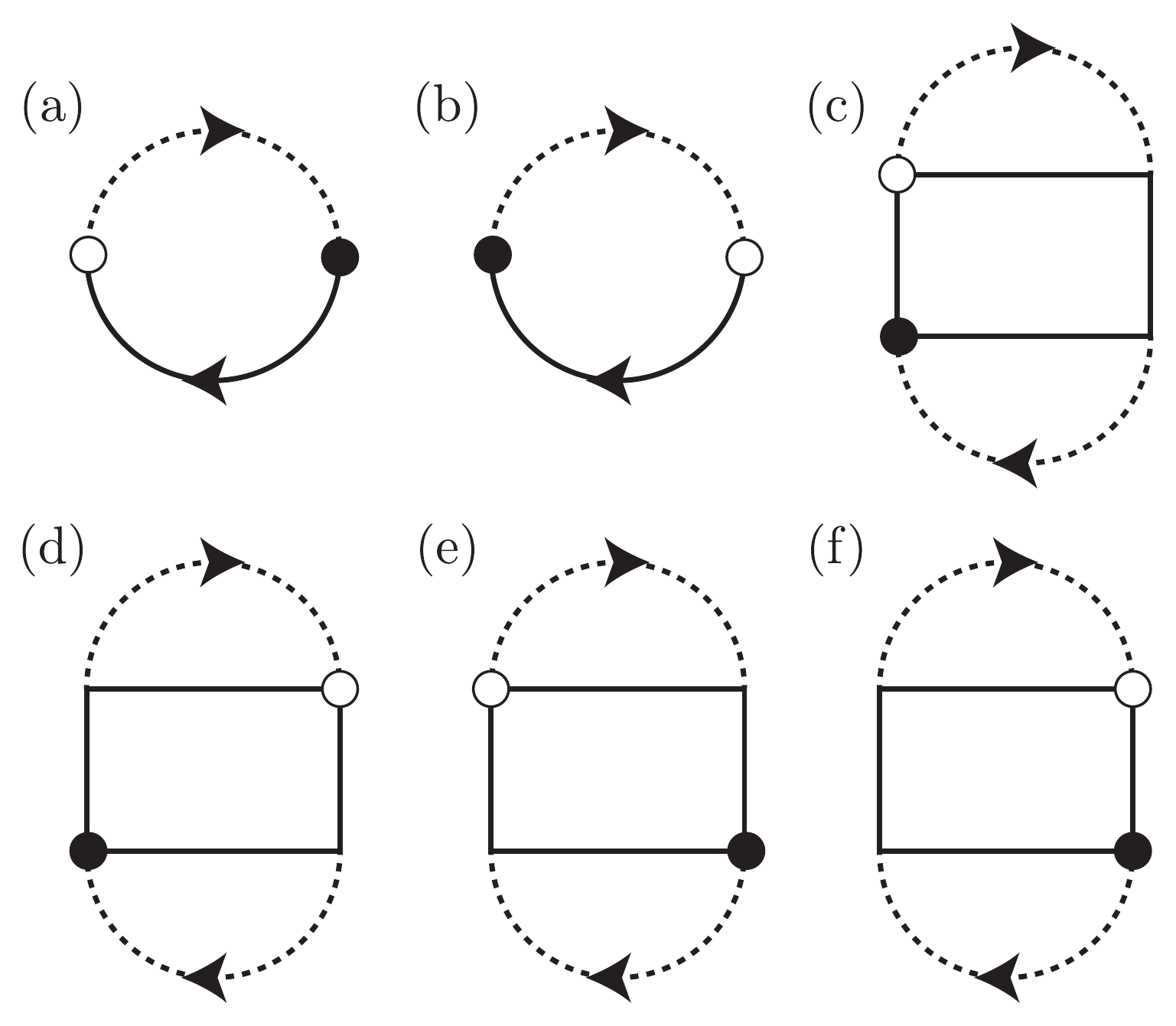}
    \caption{Feynman diagram representing $S_{r_1r_2,(a)}^{\nu_1\nu_2}(t_1,t_2), \cdots , S_{r_1r_2,(f)}^{\nu_1\nu_2}(t_1,t_2)$.
    The solid line denotes the full GF, $G_s$, and the broken line denotes the 1PI self-energy, $\Sigma_{rs}$.
    The rectangle denotes the two-particle GF, $D_{s_1s_2}$.
    The open dot denotes $t_1$ and the filled dot denotes $t_2$.
    }
    \label{fig:current_diagram}
\end{figure}

The charge current noise for the symmetric coupling case is defined as
\begin{eqnarray}
    S = \int dt_1 \Braket{\delta I(t_1) \delta I(0)} , \label{eqn:noise_definition_general}
\end{eqnarray}
where $\delta I (t) = I(t) - \Braket{I(t)}$.
This quantity is calculated in the Keldysh formalism as follows:
\begin{eqnarray}
    S = \frac{1}{2} \int dt_1 \ S^{+-}(t_1,0) + S^{-+}(t_1,0) ,
\end{eqnarray}
where
\begin{eqnarray}
    S^{\nu_1\nu_2}(t_1,t_2) &=& \frac{1}{4} (S_{\mathrm{L}\mathrm{L}}^{\nu_1\nu_2}(t_1,t_2) + S_{\mathrm{R}\mathrm{R}}^{\nu_1\nu_2}(t_1,t_2)  \nonumber \\
    && \hspace{1.0cm} - S_{\mathrm{L}\mathrm{R}}^{\nu_1\nu_2}(t_1,t_2) - S_{\mathrm{R}\mathrm{L}}^{\nu_1\nu_2}(t_1,t_2) ) , \nonumber \\
\end{eqnarray}
\begin{eqnarray}
    S_{r_1r_2}^{\nu_1\nu_2}(t_1,t_2) &= \Braket{\mathcal{T}_K \delta I_{r_1}(t_{1,\nu_1}) \delta I (t_{2,\nu_2})} .
\end{eqnarray}
Using Wick's theorem, $S_{r_1r_2}^{\nu_1\nu_2}(t_1,t_2)$ is calculated by the summation of six diagrams
\begin{eqnarray}
    && S_{r_1r_2}^{\nu_1\nu_2}(t_1,t_2) \nonumber \\
    &&= [ (S_{r_1r_2,(a)}^{\nu_1\nu_2}(t_1,t_2) + S_{r_1r_2,(b)}^{\nu_1\nu_2}(t_1,t_2)) \delta_{r_1,r_2} \nonumber \\
    &&\hspace{1.0cm} + S_{r_1r_2,(c)}^{\nu_1\nu_2}(t_1,t_2) + S_{r_1r_2,(d)}^{\nu_1\nu_2}(t_1,t_2) \nonumber \\
    &&\hspace{1.0cm} + S_{r_1r_2,(e)}^{\nu_1\nu_2}(t_1,t_2) + S_{r_1r_2,(f)}^{\nu_1\nu_2}(t_1,t_2) ] . 
\end{eqnarray}
The Feynman diagram for each term is shown in Fig. \ref{fig:current_diagram}.
Summing all the diagram contributions, one obtains Eq. (\ref{eqn:current_noise_general}).
\begin{eqnarray}
    S &=& \frac{e^2}{4} \sum_{s} \int \frac{d\omega}{2\pi} \  i\frac{\Gamma}{2} [  G_s^{-+}(\omega) f_{+}(\omega) +  G_s^{+-}(\omega) (f_{+}(\omega) - 2)] \nonumber \\
    && - \frac{e^2}{4} \sum_{s_1,s_2} \int \frac{d\omega_1 d\omega_2}{(2\pi)^2} \ \frac{\Gamma^2}{4} f_{-}(\omega_1) f_{-}(\omega_2) D_{s_1s_2}(\omega_1,\omega_2) . \nonumber \\ \label{eqn:current_noise_general}
\end{eqnarray}
A detailed derivation of Eq. (\ref{eqn:current_noise_general}) is given in Appendix \ref{apx:ChargeCurrentNoiseGeneral}.
Here $f_+(\omega)$ and $f_-(\omega)$ are defined as
\begin{eqnarray}
    f_-(\omega) &=& f_{\mathrm{L}}(\omega)-f_{\mathrm{R}}(\omega) , \\
    f_+(\omega) &=& f_{\mathrm{L}}(\omega)+f_{\mathrm{R}}(\omega) ,
\end{eqnarray}
and $D_{s_1s_2}(\omega_1,\omega_2)$ is defined as 
\begin{eqnarray}
    D_{s_1s_2}(\omega_1,\omega_2) &=& D_{s_1s_2}^{+-+-}(\omega_1,\omega_2) - D_{s_1s_2}^{+--+}(\omega_1,\omega_2)  \nonumber \\
    &&\hspace{0.5cm} - D_{s_1s_2}^{-++-}(\omega_1,\omega_2) +  D_{s_1s_2}^{-+-+}(\omega_1,\omega_2) ,  \nonumber \\ \label{eqn:definition_Dfunction}
\end{eqnarray}
where $D_{s_1s_2}^{\nu_1\nu_2\nu_3\nu_4}(\omega_1,\omega_2)$ is the special case of the Fourier component
\begin{eqnarray}
    &&D_{s_1s_2}^{\nu_1\nu_2\nu_3\nu_4}(\omega_1,\omega_2) \nonumber \\
    &&= \int dt_1 dt_2 dt_3 \  D_{s_1s_2}^{\nu_1\nu_2\nu_3\nu_4}(t_1,t_2,t_3,0) e^{i\omega_1(t_1-t_2)} e^{i\omega_2 t_3} . \nonumber \\
\end{eqnarray}

\section{Normalization conditions \label{apx:NormalizationConditions}}
The counter-terms are defined by the normalization equation shown in Eqs. (\ref{eqn:normalization_cond_1}-\ref{eqn:normalization_cond_1}).
In practical calculations, it is useful to use the equivalent equations as follows:
\begin{eqnarray}
    \lim_{\mu_{\mathrm{L}}=\mu_{\mathrm{R}}=0,T_{\mathrm{L}}=T_{\mathrm{R}}=0} \tilde{\Sigma}_s^A(0) &=& 0 , \\
    \lim_{\mu_{\mathrm{L}}=\mu_{\mathrm{R}}=0,T_{\mathrm{L}}=T_{\mathrm{R}}=0} \frac{\partial}{\partial \omega} \left. \tilde{\Sigma}_s^A(\omega)\right|_{\omega=0} &=& 0 , \\
    \lim_{\mu_{\mathrm{L}}=\mu_{\mathrm{R}}=0,T_{\mathrm{L}}=T_{\mathrm{R}}=0} \tilde{\Gamma}_{s_1s_2}^{++++}(0,0,0,0) &=& i \tilde{U} \delta_{s_1,\bar{s}_2} .
\end{eqnarray}

\section{Derivation of noise in RPT \label{apx:DerivationNoiseRPT}}
In this appendix, we discuss the second-order perturbation of Eq. (\ref{eqn:current_noise_general}) with respect to $\tilde{U}$.
Throughout this appendix, we drop the spin index in the full GFs and 1PI self-energy because they are spin-independent.
In addition, we define the non-interacting GFs as
\begin{eqnarray}
    G_0^{\nu_1 \nu_2}(\omega) = \left. G^{\nu_1\nu_2}(\omega) \right|_{\tilde{U}=0} .
\end{eqnarray}

First, we consider the non-vertex part (the former term in Eq. (\ref{eqn:current_noise_general})):
\begin{eqnarray}
    S_{\mathrm{nv}} &=& \frac{e^2}{4} \sum_{s} \int \frac{d\omega}{2\pi} \  i\frac{\tilde{\Gamma}}{2} [  G^{-+}(\omega) f_{+}(\omega) \nonumber \\
    && \hspace{3.5cm} +  G^{+-}(\omega) (f_{+}(\omega) - 2)]   \nonumber \\
    &&+ \frac{e^2}{4} \sum_{s} \int \frac{d\omega}{2\pi}  \ \frac{\tilde{\Gamma}^2}{4} \{ [G^A(\omega)]^2 +  [G^R(\omega)]^2 \} f_{-}^2(\omega) \nonumber \\
    &=& \frac{e^2}{2\pi} \frac{\tilde{\Gamma}^2}{4} \int d\omega \  G^R(\omega)G^A(\omega)  \Bigl[ 2f_{\mathrm{L}}(\omega)  (1-f_{\mathrm{L}}(\omega)) \nonumber \\
    && \hspace{3.0cm} + 2 f_{\mathrm{R}}(\omega) (1-f_{\mathrm{R}}(\omega)) + f_{-}^{2}(\omega) \Bigr]   \nonumber \\
    && + \frac{1}{2} \frac{e^2}{2\pi}  \int d\omega \  i\frac{\tilde{\Gamma}}{2} G^R(\omega) G^A(\omega) \Bigl[  \tilde{\Sigma}^{-+}(\omega) f_{+}(\omega)  \nonumber \\
    && \hspace{4.0cm} + \tilde{\Sigma}^{+-}(\omega) (f_{+}(\omega) - 2) \Bigr]  \nonumber \\
    && + \frac{1}{2} \frac{e^2}{2\pi} \frac{\tilde{\Gamma}^2}{4} \int d\omega  \  \{ [G^A(\omega)]^2 +  [G^R(\omega)]^2 \} f_{-}^2(\omega) . \nonumber \\
\end{eqnarray}
For the electrochemical potential bias case, each term is evaluated up to $(\Delta \mu)^3$ as
\begin{eqnarray}
    &&\frac{e^2}{2\pi} \frac{\tilde{\Gamma}^2}{4} \int d\omega \  G^R(\omega)G^A(\omega)    f_{-}^{2}(\omega)    \nonumber \\
    &&= \frac{e^2}{2\pi} \frac{\tilde{\Gamma}^2}{4} \Biggl[ \Bigl\{ G_0^R(0) G_0^A(0) + 2 \mathrm{Re} \left[ G_0^R(0) [G_0^A(0)]^2 \tilde{\Sigma}^A(0) \right] \Bigr\}   \Delta \mu  \nonumber \\
    &&\hspace{1.5cm} + \frac{1}{24} \Biggl\{ \left. \frac{\partial^2}{\partial \omega^2} [G_0^R(\omega) G_0^A(\omega)] \right|_{\omega=0}  \nonumber \\
    &&\hspace{2.5cm} + 2 \mathrm{Re} \left[ G_0^R(\omega) [G_0^A(\omega)]^2   \frac{\partial^2
\tilde{\Sigma}^A(\omega)}{\partial \omega^2} \right]_{\omega=0} \Biggr\} (\Delta \mu)^3 \Biggr]  \nonumber \\
    && \hspace{5.0cm} + O((\Delta \mu)^4) , \label{eqn:nvterm_1}
\end{eqnarray}
\begin{eqnarray}
    &&\frac{e^2}{2\pi} \frac{\tilde{\Gamma}^2}{4} \int d\omega \  \{ [G^A(\omega)]^2 +  [G^R(\omega)]^2 \}  f_{-}^{2}(\omega)    \nonumber \\
    &&= \frac{e^2}{2\pi} \frac{\tilde{\Gamma}^2}{4} \Biggl[ \left\{ 2\mathrm{Re}\left[ [G_0^A(0)]^2 + 2[G_0^A(0)]^3 \tilde{\Sigma}^A(0) \right] \right\}   \Delta \mu  \nonumber \\
    &&\hspace{1.0cm} + \frac{1}{24}  2 \mathrm{Re} \left[  6[G_0^A(\omega)]^4 + 2 [G_0^A(\omega)]^3   \frac{\partial^2
\tilde{\Sigma}^A(\omega)}{\partial \omega^2} \right]_{\omega=0} (\Delta \mu)^3 \Biggr] \nonumber \\
    &&\hspace{5.0cm} + O((\Delta \mu)^4) , \label{eqn:nvterm_2}
\end{eqnarray}
\begin{eqnarray}
    &&\frac{1}{2} \frac{e^2}{2\pi}  \int d\omega \  i\frac{\tilde{\Gamma}}{2} G^R(\omega) G^A(\omega)  [  \tilde{\Sigma}^{-+}(\omega) f_{+}(\omega) + \tilde{\Sigma}^{+-}(\omega) (f_{+}(\omega) - 2) ]  \nonumber \\
    &&= \frac{e^2}{2\pi} \frac{\tilde{U}^2}{(2\pi)^2} \int d\omega d\omega_1 d\omega_2  \  \Bigl[ \tilde{A}(\omega_1) \tilde{A}(\omega_2) \tilde{A}(\omega) \tilde{A}(\omega-\omega_1+\omega_2)  \nonumber \\
    &&\hspace{2.0cm} \times f_+(\omega_1) f_+(-\omega_2) f_+(\omega)  f_+(-\omega+\omega_1-\omega_2) \Bigr]  ,
\end{eqnarray}
where $\tilde{A}(\omega)$ is the spectrum function with the renormalized parameters defined as
\begin{eqnarray}
    \tilde{A}(\omega) = \frac{\tilde{\Gamma}/2}{(\omega-\tilde{\epsilon}_d)^2+\tilde{\Gamma}^2/4} .
\end{eqnarray}
The 1PI self-energy is calculated exactly up to the $\omega^2$,$(\Delta \mu)^2$, and $(\Delta T)^2$ order within the second-order perturbation for the particle-hole symmetric case~\footnote{One might think that the coefficients of $(\Delta T)^2$ is different from the result in Ref. \cite{Oguri2001Sep,Sakano2011Feb}. One can obtain Eq. (\ref{eqn:1PIself_adv}) by replacing, $T^2 = (T_L^2 + T_R^2) / 2$.}:
\begin{eqnarray}
    \tilde{\Sigma}^A(\omega) &=& i \frac{\tilde{U}^2}{(2\pi)^2} \frac{8}{\tilde{\Gamma}^2} \left[ 2 \omega^2 + \frac{3}{2} (\Delta \mu)^2 + 6 \zeta(2) (\Delta T)^2 \right]  \nonumber \\
    &&\hspace{2.0cm} + O(\omega^3,(\Delta \mu)^3,(\Delta T)^3) . \label{eqn:1PIself_adv}
\end{eqnarray}
Substituting Eq. (\ref{eqn:1PIself_adv}) into Eqs. (\ref{eqn:nvterm_1}) and (\ref{eqn:nvterm_2}) and using the following equation
\begin{eqnarray}
    &&\int dx dy dz \ F(x,y,z) f_{r_1}(x) f_{r_2}(-y) f_{r_3}(z) f_{r_4}(-x+y-z) \nonumber \\
    &&= \frac{1}{6} F(0,0,0) |\mu_{r_1} - \mu_{r_2} + \mu_{r_2} - \mu_{r_4}|^3  + O((\Delta \mu)^4) , 
\end{eqnarray}
one obtains the non-vertex part as
\begin{eqnarray}
    S_{\mathrm{nv},\Delta \mu} = \frac{e^2}{2\pi} \frac{4}{\tilde{\Gamma}^2} (\Delta \mu)^3 \left[ \frac{1}{6}  + \frac{1}{2} (R-1)^2 \right] .
\end{eqnarray}
For the temperature bias case, the non-vertex part is calculated in the same way:
\begin{eqnarray}
    &&\frac{e^2}{2\pi} \frac{\tilde{\Gamma}^2}{4} \int d\omega \  G^R(\omega)G^A(\omega)    f_{-}^{2}(\omega)    \nonumber \\
    &&= \frac{e^2}{2\pi} \frac{\tilde{\Gamma}^2}{4} \Biggl\{ (2\ln 2 -1) \mathrm{Re} \Bigl[ G_0^R(0) G_0^A(0)  \nonumber \\
    && \hspace{4.0cm} + 2 G_0^R(0) [G_0^A(0)]^2 \tilde{\Sigma}^A(0) \Bigr]    \Delta T  \nonumber \\
    &&\hspace{1.5cm} + \left( \frac{3}{2} \zeta(3) - \zeta(2) \right)  \mathrm{Re} \Biggl[  \frac{\partial^2}{\partial \omega^2} [G_0^R(\omega) G_0^A(\omega)] \nonumber \\
    &&\hspace{2.5cm} + 2  G_0^R(\omega) [G_0^A(\omega)]^2   \frac{\partial^2
\tilde{\Sigma}^A(\omega)}{\partial \omega^2} \Biggr]_{\omega=0}  (\Delta T)^3 \Biggr\} \nonumber \\
    &&\hspace{4.0cm}  + O((\Delta T)^4) , \label{eqn:nvterm_1_temp}
\end{eqnarray}
\begin{eqnarray}
    &&\frac{e^2}{2\pi} \frac{\tilde{\Gamma}^2}{4} \int d\omega \  \{ [G^A(\omega)]^2 +  [G^R(\omega)]^2 \}  f_{-}^{2}(\omega)    \nonumber \\
    &&= \frac{e^2}{2\pi} \frac{\tilde{\Gamma}^2}{4} \Biggl\{ 2(2\ln 2 -1)  \mathrm{Re}\Bigl[ [G_0^A(0)]^2  + 2[G_0^A(0)]^3 \tilde{\Sigma}^A(0) \Bigr]    \Delta T  \nonumber \\
    &&\hspace{1.5cm} + 2 \left( \frac{3}{2} \zeta(3) - \zeta(2) \right)  \mathrm{Re} \Biggl[  6[G_0^A(\omega)]^4 \nonumber \\
    &&\hspace{3.5cm} + 2 [G_0^A(\omega)]^3   \frac{\partial^2
\tilde{\Sigma}^A(\omega)}{\partial \omega^2} \Biggr]_{\omega=0} (\Delta T)^3 \Biggr\} \nonumber \\
    &&\hspace{5.0cm} + O((\Delta T)^4) , \label{eqn:nvterm_2_temp}
\end{eqnarray}
\begin{eqnarray}
    &&2 \frac{e^2}{2\pi} \frac{\tilde{\Gamma}^2}{4} \int d\omega \  G^R(\omega)G^A(\omega)  f_{\mathrm{L}}(\omega)  (1-f_{\mathrm{L}}(\omega))    \nonumber \\
    &&= 2 \frac{e^2}{2\pi} \frac{\tilde{\Gamma}^2}{4} \Biggl\{ \mathrm{Re} \left[ G_0^R(0) G_0^A(0) + 2 G_0^R(0) [G_0^A(0)]^2 \tilde{\Sigma}^A(0) \right]    \Delta T  \nonumber \\
    &&\hspace{2.0cm} + \zeta(2) \mathrm{Re} \Biggl[  \frac{\partial^2}{\partial \omega^2} [G_0^R(\omega) G_0^A(\omega)] \nonumber \\
    &&\hspace{3.0cm} + 2  G_0^R(\omega) [G_0^A(\omega)]^2   \frac{\partial^2
\tilde{\Sigma}^A(\omega)}{\partial \omega^2} \Biggr]_{\omega=0}  (\Delta T)^3 \Biggr\} \nonumber \\
    &&\hspace{5.0cm} + O((\Delta T)^4) . \label{eqn:nvterm_3_temp}
\end{eqnarray}
Substituting Eq. (\ref{eqn:1PIself_adv}) into Eqs. (\ref{eqn:nvterm_1_temp} - \ref{eqn:nvterm_3_temp}) and using the following equation:
\begin{eqnarray}
    &&\int dx dy dz \ F(x,y,z) f_{r_1}(x) f_{r_2}(-y) f_{r_3}(z) f_{r_4}(-x+y-z) \nonumber \\
    &&= \left( \frac{(2 + 6 \ln 2) \pi^2}{3} + 18 \zeta(3) \right) F(0,0,0) (\Delta T)^3 + O((\Delta T)^4) , \nonumber \\
\end{eqnarray}
the non-vertex part for the temperature bias case is calculated as
\begin{eqnarray}
    S_{\mathrm{nv},\Delta T} &=& \frac{e^2}{\pi} \Delta T + \frac{e^2}{2\pi} \frac{4}{\tilde{\Gamma}^2} (\Delta T)^3 \Biggl[ (6\zeta(3)-8\zeta(2)) \nonumber \\
    &&+\left( \frac{9}{2} \zeta(3) + (3 \ln 2 - 9) \zeta(2) \right) (R-1)^2 \Biggr] \nonumber \\
    &&\hspace{3.5cm} + O((\Delta T)^4).
\end{eqnarray}

Next, we consider the vertex part (the latter term in Eq. (\ref{eqn:current_noise_general})):
\begin{eqnarray}
    S_{\mathrm{v}} 
    &=& - \frac{e^2}{4} \sum_{s_1,s_2} \int \frac{d\omega_1 d\omega_2}{(2\pi)^2} \ \frac{\tilde{\Gamma}^2}{4} f_{-}(\omega_1) f_{-}(\omega_2) D_{s_1s_2}(\omega_1,\omega_2) \nonumber \\
    &=& - \frac{e^2}{(2\pi)^2} \int d\omega_1 d\omega_2 \ \Biggl\{ \frac{\Gamma^2}{4} f_{-}(\omega_1) f_{-}(\omega_2)  \nonumber \\
    &&\hspace{1.0cm} \times \mathrm{Re} \Bigl[  [G^R(\omega)]^2 [G^R(\omega_2)]^2  \tilde{\Gamma}_{a}(\omega_1,\omega_2)   \nonumber \\
    &&\hspace{2.0cm} - [G^R(\omega_1)]^2 [G^A(\omega_2)]^2  \tilde{\Gamma}_{b}(\omega_1,\omega_2) \nonumber \\
    &&\hspace{2.0cm} + 2  \left( G^{--}(\omega_2) G^{++}(\omega_2) - [G^{+-}(\omega_2)]^2 \right) \nonumber \\
    &&\hspace{2.5cm} \times [G^R(\omega_1)]^2 \tilde{\Gamma}_c(\omega_1,\omega_2) \Bigr] \Biggr\}  , \label{eqn:vertex_part}
\end{eqnarray}
where $\tilde{\Gamma}_{a}(\omega_1,\omega_2)$, $\tilde{\Gamma}_{a}(\omega_1,\omega_2)$, and $\tilde{\Gamma}_{a}(\omega_1,\omega_2)$ are the specific component of the four-point full vertex defined as 
\begin{eqnarray}
    \tilde{\Gamma}_{a}(\omega_1,\omega_2) &=& \frac{1}{2} \sum_{s_1,s_2} \sum_{\nu_1,\nu_2} \tilde{\Gamma}_{s_1,s_2}^{+\nu_1+\nu_2}(\omega_1,\omega_2) , \\
    \tilde{\Gamma}_{b}(\omega_1,\omega_2) &=& \frac{1}{2} \sum_{s_1,s_2} \sum_{\nu_1,\nu_2} \tilde{\Gamma}_{s_1,s_2}^{+\nu_1\nu_2-}(\omega_1,\omega_2) ,  \\
    \tilde{\Gamma}_{c}(\omega_1,\omega_2) &=& \frac{1}{2} \sum_{s_1,s_2} \sum_{\nu_1,\nu_2,\nu_3} \tilde{\Gamma}_{s_1,s_2}^{+\nu_1\nu_2\nu_3}(\omega_1,\omega_2) .
\end{eqnarray}
The four-point full vertex function is calculated exactly up to the $\omega_1$, $\omega_2$, $\Delta \mu$, and $\Delta T$ order within the second-order perturbation for the particle-hole symmetric case:
\begin{eqnarray}
    &&\tilde{\Gamma}_{a}(\omega_1,\omega_2) \nonumber \\
    &&= i \tilde{U} - \frac{\tilde{U}^2}{2\pi} \frac{4}{\tilde{\Gamma}^2} \left[ 2(-\omega_1 + 3 \omega_2) + \Delta \mu + (1+2\ln2) \Delta T \right] \nonumber \\
    &&\hspace{0.5cm} - 2 \frac{\tilde{U}^2}{2\pi} \int d\omega \ G_0^{+-}(\omega) G_0^{-+}(\omega-\omega_1 +\omega_2) \nonumber \\
    &&\hspace{3.0cm} + O(\omega_1^2,\omega_2^2,(\Delta T)^2,(\Delta \mu)^2 ) , \label{eqn:vertex_a}
\end{eqnarray}
\begin{eqnarray}
    &&\tilde{\Gamma}_{b}(\omega_1,\omega_2) \nonumber \\
    &&= \frac{\tilde{U}^2}{2\pi} \frac{4}{\tilde{\Gamma}^2} \left[\Delta \mu + (1+2\ln2) \Delta T \right] \nonumber \\
    &&\hspace{0.5cm} - \frac{\tilde{U}^2}{2\pi} \int d\omega \ G_0^{+-}(\omega) G_0^{+-}(\omega_1 +\omega_2 -\omega) \nonumber \\
    &&\hspace{3.0cm} + O(\omega_1^2,\omega_2^2,(\Delta T)^2,(\Delta \mu)^2 ) , \label{eqn:vertex_b}
\end{eqnarray}
\begin{eqnarray}
    &&\tilde{\Gamma}_{c}(\omega_1,\omega_2) = i \tilde{U} - \frac{\tilde{U}^2}{2\pi} \frac{4}{\tilde{\Gamma}^2}  2(-\omega_1 + 3 \omega_2)  \nonumber \\
    &&\hspace{3.0cm} + O(\omega_1^2,\omega_2^2,(\Delta T)^2,(\Delta \mu)^2 ) .  \label{eqn:vertex_c}
\end{eqnarray}
Substituting Eqs. (\ref{eqn:vertex_a}-\ref{eqn:vertex_c}) into Eq. (\ref{eqn:vertex_part}) and using the following equations for the integrals of the Fermi distribution functions:
\begin{eqnarray}
    &&\int dx dy dz \ F(x,y,z) f_-(x) f_-(y) f_+(z) (f_+(z-x+y) - 2) \nonumber \\
    &&= F(0,0,0) \frac{4}{3} (\Delta \mu)^3 + O((\Delta \mu)^4) ,
\end{eqnarray}
\begin{eqnarray}
    &&\int dx dy dz \ F(x,y,z) f_-(x) f_-(y) f_+(z) f_+(x+y-z)  \nonumber \\
    &&= F(0,0,0) \frac{4}{3} (\Delta \mu)^3 + O((\Delta \mu)^4) ,
\end{eqnarray}
\begin{eqnarray}
    &&\int dx dy dz \ F(x,y,z) f_-(x) f_-(y) f_+(z) (f_+(z-x+y) - 2) \nonumber \\
    &&= F(0,0,0) \left( 4 \zeta(3) - \frac{2\pi^2}{3} \right) (\Delta T)^3 + O((\Delta T)^4) ,
\end{eqnarray}
\begin{eqnarray}
    &&\int dx dy dz \ F(x,y,z) f_-(x) f_-(y) f_+(z) f_+(x+y-z)  \nonumber \\
    &&= F(0,0,0) \left( 4 \zeta(3) - \frac{2\pi^2}{3} \right) (\Delta T)^3 + O((\Delta T)^4) ,
\end{eqnarray}
one obtains the vertex part in the electochemical potential bias case as
\begin{eqnarray}
    S_{\mathrm{v}, \Delta \mu} = \frac{e^2}{2\pi} \frac{4}{\tilde{\Gamma}^2} (\Delta \mu)^3 (R-1)^2  + O((\Delta \mu)^4),
\end{eqnarray}
and that in the temperature bias case as
\begin{eqnarray}
    S_{\mathrm{v},\Delta T} &=& \frac{e^2}{2\pi} \frac{4}{\tilde{\Gamma}^2} (\Delta T)^3 (R-1)^2 \left( \zeta(3) - 3 \zeta(2) \right)  \nonumber \\
    &&\hspace{3.0cm} + O((\Delta T)^4). 
\end{eqnarray}

As a result, the total noise in the electochemical potential bias case is calculated as
\begin{eqnarray}
    S &=& S_{\mathrm{v},\Delta \mu} + S_{\mathrm{nv},\Delta \mu} \nonumber \\
    &=& \frac{2e^2}{\pi} \frac{(\Delta \mu)^3}{\tilde{\Gamma}^2}\left[ \frac{1}{6} + \frac{3}{2} (R-1)^2 \right] + O((\Delta \mu)^4) , \nonumber \\
\end{eqnarray}
and that in the temperature bias case as
\begin{eqnarray}
    S &=& S_{\mathrm{v},\Delta T} + S_{\mathrm{nv},,\Delta T} \nonumber \\
    &=& \frac{e^2}{\pi} \Delta T + \frac{2e^2}{\pi}\frac{(\Delta T)^3}{\tilde{\Gamma}^2} \Biggl[ (6\zeta(3)-8\zeta(2))  \nonumber \\
    &&\hspace{1.5cm} + \left( \left( 3 \ln 2  - 12 \right) \zeta(2) + \frac{11}{2} \zeta(3) \right) (R-1)^2  \Biggr] \nonumber \\
    &&\hspace{4.0cm} + O((\Delta T)^4) . \label{eqn:totalnoise}
\end{eqnarray}
To obtain the noise $S_{\ell}$, it is necessary to subtract the equilibrium noise, which is calculated in the case, $T_{\mathrm{L}}=T_{\mathrm{R}}=\Delta T$.
The equilibrium noise can be calculated by Eq. (\ref{eqn:nvterm_3_temp}) as
\begin{eqnarray}
    &&S(\Delta T,\Delta T) = \frac{4e^2}{\pi} \Delta T - \frac{4e^2}{\pi^2}\frac{(\Delta T)^3}{\tilde{\Gamma}^2} \left[ 4 \zeta(2) + 8 \zeta(2)  (R-1)^2 \right]  \nonumber \\
    &&\hspace{4.0cm} + O((\Delta T)^4) .  \label{eqn:eqnoise}
\end{eqnarray}
Subtracting the equilibrium noise (Eq. (\ref{eqn:eqnoise})) from the total noise (Eq. (\ref{eqn:totalnoise})), one obtains Eq. (\label{eqn:curent_noise_RPT_temp}-\ref{eqn:curent_noise_coef_RPT_temp}).

\bibliography{references}

\end{document}